# Electrostatic Climber for Space Elevator and Launcher*

**Alexander Bolonkin**
*C&R, 1310 Avenue R, #6-F, Brooklyn, NY 11229, USA.*
*<aBolonkin@juno.com>,< aBolonkin@gmail.com>, <http://Bolonkin.narod.ru>*

**Abstract**

  Author details research on the new, very prospective, electrostatic Space Elevator climber based on a new electrostatic linear engine  previously offered at the 42nd Joint Propulsion Conference (AIAA-2006-5229) and published in AEAT, Vol.78, No.6, 2006, pp. 502-508. The electrostatic climber discussed can have any speed (and braking), the energy for climber movement is delivered by a lightweight high-voltage line into a Space Elevator-holding cable from Earth electric generator. This electric line also can be used for delivery electric energy to a Geosynchronous Space Station. At present, the best solution of the climber problem (announced by NASA as one important awarding problem of Space Elevator) is problematic.

   Author also shows the linear electrostatic engine may be used as realistic power space launcher at the present time. Two projects illustrate these new devices.

  *This work is presented as paper AIAA-2007-5838 for 43 Joint Propulsion Conference, Cincinnati, Ohio, USA, 9-11 July, 2007,

## Introduction

  **General**.  The aviation, space, and energy industries need revolutionary ideas which will significantly improve the capability of future ground, air and space vehicles. The author has offered a series of new ideas [1-73] contained in (a) numerous patent applications [3 -17], in (b) manuscripts that have been presented at the World Space Congress (WSC)-1992, 1994 [19 -22], the WSC-2002 ]23 -31], and numerous Propulsion Conferences [32 -39], and (c) other articles [40 -73].

     In this article a revolutionary method and implementations for future space flights and ground systems are proposed. The method uses highly charged cylindrical bodies. The proposed space launch system creates tens of tons of thrust and accelerates space apparatus to high speeds.

   **History.** In early works and patent applications (1965 - 1991), in World Space Congress-2002 and other scientific forums the author suggested a series new cable launchers, space transport systems, space elevator, anti-gravitator, kinetic space tower, and other systems, which decrease the cost of space launch in thousands  times or increase the possibilities ground systems. All of them need in linear engine. In particular, there are: Cable Space Launcher [23-25, 40], Earth-Moon Transport system [29,39], Earth-Mars Transport System [30], Circle Space Launcher [31], Hypersonic tube gas launcher [32], Air Cable aircraft [41, 42], Non-Rocket Transport System for Space Elevator (Elevator climber)[36], Centrifugal Keeper [38], Asteroid Propulsion System [27, 40], Kinetic Space Towers [43],  Long Transfer of Mechanical Energy [45], High Speed Catapult Aviation [52], Kinetic Anti-Gravitator [55], Electrostatic Levitation [59], AB Levitator [67] and so on [1]-[73] (Fig.1). Part of these works is in books [60],[73].

  The author offered an electrostatic engine [66] which can be used for every noted installation as driver. In this article author considers more detail the application of offered electrostatic linear engine to the space elevator climber and to Earth space launcher (accelerator) of spaceship and probes.

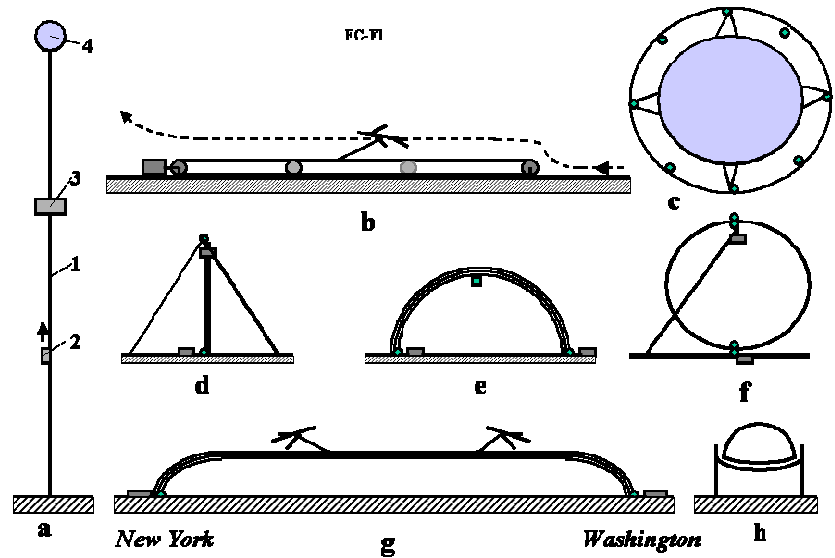

**Fig.1**. Installations needing the linear electrostatic engine. (a) Space elevator [36]. Notation: 1 - space elevator, 2 - climber, 3 - geosynchronous space station, 4 - balancer of space elevator. (b) Space cable launcher [23-25, 40]; (b) Circle launcher [31]; (c) Earth round cable space keeper [38]; (d) Kinetic space tower [43]; (e - f) Space keeper [38]; (g) (f) Cable aviation [41,42]; (g) Levitation train [59].

## Description of Electrostatic Linear Climber and Launcher

   The linear electrostatic engine [66] (climber) for Space Elevator includes the following main parts (fig.2): plate (type) stator 1 (special cable of Space elevator), cylinders 3 inside having conducting layer (or net) (cylinder may be vacuum or inflatable film), insulator of conducting layer, chargers (switches 6) of cable cylinders, high voltage electric current line 6, linear rotor 7. Liner rotor has permanent charged cylinder 4. As additional devices the engine can have a gas compaction, and vacuum pump [66].

   The cable (stator) has a strong cover 2 (it keeps tensile stress - thrust/braking) and variable cylindrical charges contained dielectric cover (insulator). The conducting layer is very thin and we neglect its weight. Cylinders of film are also very lightweight. The charges can be connected to high-voltage electric lines 6 linked to high-voltage device (electric generator) located on the ground.

   The electrostatic engine works in the following mode. The rotor has a stationary positive charge. The cable has the variable positive and negative charges. These charges can be received by connection to the positive or negative high voltage electric line located in cable (in stator). When positive rotor charge is located over given stator cylinder this cylinder connected by switch to positive electric lines and cylinder is charged positive charge but simultaneously the next stator cylinder is charged by negative charge. As result the permanent positive rotor charge repels from given positive stator charge and attract to the next negative stator charge. This force moves linear rotor (driver). When positive rotor charge reaches a position over the negative stator charge that charge re-charges to positive charge and next cylinder is connected to the negative electric line and the cycle is repeated. For increasing the efficiency, the positive and negative stator charges before the next cycle can run down through special device and their energy is returned to the electric line. Linear electrostatic engine can have very high efficiency.

Earth constant potential generator creates a running single wave of the charges along stationary stator. This wave (charges) attracts (repel) the opposed (same) charges in rotor (linear driver) and moves (thrust or brake use) climber.

The space launcher works same (fig. 2d, 2e). That has a stationary stator and mobile rotor (driver). The stationary stator (monorail) located along the Earth's surface. Driver is connected to space-aircraft and accelerates the aircraft to a needed speed (8 km/s and more) [23-25]. For increasing a thrust the driver of the space launcher can have some charges (fig. 2d) separated by enough neutral non charged stator cylinders.

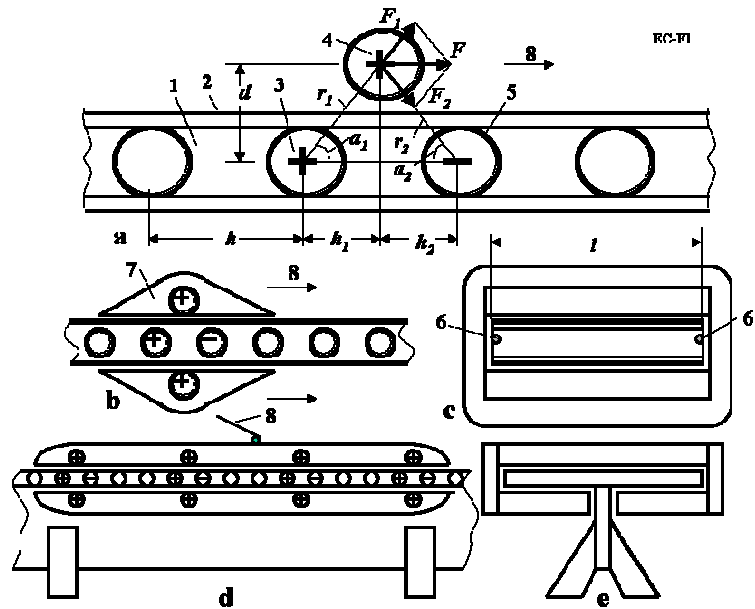

**Fig. 2.** Electrostatic linear engine (accelerator) for Space Elevator and Space Launcher [23 -25, 40]. (a) Explanation of force in electrostatic engine; (b) Two cylindrical electrostatic engine for Space Elevator (side view); (c) Two cylindrical electrostatic engine for Space Elevator (forward view); (d) Eight cylindrical electrostatic engine for Space Launcher (side view); (e) Eight cylindrical electrostatic engine for Space Launcher (forward view).

Notations: 1 - plate cable of Space Elevator with inserted variable cylindrical charges; 2 - part of cable bearing tensile stress; 3 - insulated variable charging cylinder of stator; 4 - insulated permanent charging cylinder of rotor; 6 – high-voltage wires connected with Earth's generator and switch; 7 - mobile part of electrostatic linear engine; 8 - cable to space aircraft.

Bottom and top parts of cable (or stator) have small different charge values. This difference creates a vertical electric field which supports the driver in suspended position about stator, non-contact bearing and zero friction. The driver position about stator is controlled by electronic devices.

Charges have cylindrical form (row of cylinders) and located inside a good dielectric having high disruptive voltage. The cylinders have conducting layer which allows changing the charges with high frequency and produces a running high-voltage wave of charges. The offered engine creates a large thrust (see computation below), reaches a very high (virtually unlimited) variable speed of driver (km/s), to change the moving of driver in opposed direction, to fix a driver in given position. The electrostatic engine can also operate as a high-voltage electric generator when climber (ship) is braking or moved by mechanical force. The Space Elevator climber (and many other mobile apparatus) has constant charge; the cable (stator) has running charge. The weight of electric wires is small because the voltage is very high (see projects).

## Theory of offered electrostatic engine (accelerator)

**1. Estimation of thrust.** Let us consider a single charged cylinder 3, 4, in cable (fig. 2a). The rotor charge 4 attracts to the opposed charges and repels from the same charges 3 located into stator. Let us compute the sum force acting to this single charge.

$$E_i = k\frac{2\tau}{\varepsilon\, r_i}, \quad F_i = \sum_i q E_i, \quad i = 1, 2, \quad q = \tau\, l, \tag{1}$$

where $E_i$ is electric intensity [V/m], $k = 9\times10^9$ is electric coefficient [N·m²/C²], $\tau$ is linear charge [C/m], $\varepsilon$ is dielectric constant, $r_i$ is distance between centers of charges [m], $F_i$ is force [N], $q$ is charge [C], $l$ is length of linear charge [m] (fig.2a).

The sum of two forces is

$$F = q(E_1\cos\alpha_1 + E_2\cos\alpha_2) = \frac{2k\tau^2 l}{\varepsilon}\left(\frac{h_1}{r_1^2} + \frac{h_2}{r_2^2}\right), \quad \text{where} \quad \cos\alpha_i = \frac{h_i}{r_i}, \quad \tau = \frac{\varepsilon a E}{2k}, \tag{2}$$

$h_i$ is horizontal distances between charges (fig. 2a), m; $a$ is radius of charged cylinders, m.

Define the relative ratios of distances (fig.2a)

$$\overline{h}_i = \frac{h_i}{a}, \quad \overline{r}_i = \frac{r_i}{a}, \quad i = 1, 2, \quad \overline{h} = \frac{h}{a}, \quad \overline{d} = \frac{d}{a}, \tag{3}$$

where $d$ is distance between center of charged cylinders of stator and rotor, m; $h$ is distance between two nearest stator charges, m.

Substitute all these values to (2) we receive

$$F = b\frac{\varepsilon a E^2 l}{k}, \quad \text{where} \quad b = \frac{1}{2}\left(\frac{\overline{h}_1}{\overline{r}_1^2} + \frac{\overline{h}_2}{\overline{r}_2^2}\right), \quad \overline{r}_i^2 = \overline{d}^2 + \overline{h}_i^2, \quad i = 1, 2, \quad \overline{h} = \overline{h}_1 + \overline{h}_2, \tag{4}$$

where $b$ is trust/brake coefficient.

If we take distance between charged rows of stator and cable $d = h = 3a$ (fig.2a), where $a$ is radius of charged cylinders, the value $b$ is:

   a) On ends of segment $h$ the coefficient $b_e = 1/12 = 0.08333$ .
   b) In the middle of the segment $h$ the coefficient is maximum $b_m = 2/15 = 0.13333$ .
   c) The average we take $b = 0.5(b_e + b_m) = 0.108$ .
For other ratio $d$, $h$ the coefficient $b$ is following: when $d = 3a$, $h = 4a$, the $b = 0.117$; when $d = h = 4a$, the $b = 0.08125$.

When we take the $N$ symmetric pair stator charges the maximum force may be estimated by equation [66]

$$F_{\max} = b_m\frac{eaE^2 l}{k}, \quad \text{where} \quad b_m = \left[\sum_{i=1}^{N}(-1)^{i+1}\frac{\overline{h}(i-0.5)}{\overline{d}^2 + (\overline{h})^2(i-0.5)^2}\right]. \tag{5}$$

For $N = i = 1$ we received equation (4). That pair has maximum force. For $N=100$, $d = 3a$, $h = 4a$, the $b_m = 0.074$, $b_e \approx 0$. $b = 0.037$,

The trust of linear electric engine can be tons from square meter of charged areas.

**2. Charge of one cylinder** is

$$q = \frac{ea^2 E l}{k}. \tag{6}$$

**3. Needed maximum voltage** is

$$U \approx E d. \tag{7}$$

**4. Needed power** is

$$P = F V, \tag{8}$$

  where $V$ is speed apparatus, m/s.

**5. Requested average electric currency** ($I$, A) is

$$I = P/U, \tag{9}$$

**6. Loss of voltage** ($\Delta U$, V) and coefficient efficiency of electric line is

$$\Delta U = IR, \quad \eta = \frac{U - \Delta U}{U}, \quad \text{where} \quad R = \rho \frac{L}{s}, \tag{10}$$

$R$ is electric resistance, $\Omega$; $\eta$ is coefficient efficiency; $\rho$ is specific electric resistance, $\Omega$·m (take from table); $L$ is length of wire, m; $s$ is cross-section area of wire, $m^2$.

  *Note*: For elevator climber the requested cross-section area is small ($<1$ $mm^2$) and the coefficient efficiency of electric line is good ($\eta > 0.5$) because the voltage is very large ($U \approx 15 \times 10^6$ V) (see Project 1).

**7. The number of main switching** is

$$\nu = 2V / h . \tag{11}$$

**8. Dielectric strength of current materials**.

  In our computation we used electric intensity over the electric strength of air $E_s \approx 3 \times 10^6$ V/m. That means the air located inside of engine between stator and cable can be ionized. That is not important because amount of air is small, we can extract (pump) the air from engine or fill up volume between stator and cable units with an appropriate dielectric liquid. We can also cover the electrode units with a thin dielectric layer having a high-voltage dielectric strength. No this problem in outer space.
  The data for computations are in Table 1.

**Table 1.** Properties of various good insulators (recalculated in metric system)

| Insulator | Resistivity Ohm-m. | Dielectric strength MV/m. $E_i$ | Dielectric constant, $\varepsilon$ | Tensile strength kg/mm², |
|---|---|---|---|---|
| Lexan | $10^{17}$–$10^{19}$ | 320–640 | 3 | 5.5 |
| Kapton H | $10^{19}$–$10^{20}$ | 120–320 | 3 | 15.2 |
| Kel-F | $10^{17}$–$10^{19}$ | 80–240 | 2–3 | 3.45 |
| Mylar | $10^{15}$–$10^{16}$ | 160–640 | 3 | 13.8 |
| Parylene | $10^{17}$–$10^{20}$ | 240–400 | 2–3 | 6.9 |
| Polyethylene | $10^{18}$–$5 \times 10^{18}$ | 40–680* | 2 | 2.8–4.1 |
| Poly (tetra-fluoraethylene) | $10^{15}$–$5 \times 10^{19}$ | 40–280** | 2 | 2.8–3.5 |
| Air (1 atm, 1 mm gap) | - | 4 | 1 | 0 |
| Vacuum ($1.3 \times 10^{-3}$ Pa, 1 mm gap) | - | 80–120 | 1 | 0 |

*For room temperature 500 – 700 MV/m. $E = 700$ MV/m for $t < 15$ C.
** 400 – 500 MV/m.
*Sources*: Encyclopedia of Science & Technology (New York, 2002, Vol. 6, p. 104, p. 229, p. 231) and Kikoin [74], p. 321.

  *Note:* Dielectric constant $\varepsilon$ can reach 4.5 - 7.5 for mica ($E$ is up 200 MV/m), 6 -10 for glasses ($E = 40$ MV/m), and 900 -3000 for special ceramics (marks are CM-1, T-900)[74] , p. 321, ($E = 13$ -28 MV/m). Ferroelectrics have $\varepsilon$ up to $10^4$ - $10^5$. Dielectric strength appreciably depends from surface roughness, thickness, purity, temperature and other conditions of materials. Very clean material without admixture (for example, quartz) can have electric strength up 1000 MV/m. It is necessary to find good isolative (insulation) materials and to research conditions which increase the dielectric strength.

**9. The half-life of the charge.** Let us estimate of lifetime of charged driver.

(a) *Charge in spherical ball.* Let us take a very complex condition; where the unlike charges are separated only by an insulator (charged spherical condenser):

$$Ri - U = 0, \quad U = \delta E, \quad E = \frac{kq}{\delta^2}, \quad R = \rho \frac{\delta}{4\pi a^2}, \quad U = \frac{q}{C}, \quad R\frac{dq}{dt} + \frac{a}{C} = 0, \quad \frac{dq}{q} = \frac{dt}{RC}, \quad C = \frac{\varepsilon a}{k},$$

$$q = q_0 \exp\left(-\frac{4\pi\varepsilon ak}{\rho\delta}t\right), \quad \frac{q}{q_0} = \frac{1}{2}, \quad -\frac{4\pi\varepsilon ak}{\rho\delta}t_h = \ln\frac{1}{2} = -0.693 \approx -0.7, \quad \text{final} \quad t_h = 0.693\frac{\rho\delta}{4\pi\varepsilon ka},$$

(12 -13)

where: $t_h$ – half-life time, [sec]; $R$ – insulator resistance, [Ohm]; $i$ – electric current, [A]; $U$ – voltage, [V]; $\delta$ – thickness of insulator, [m]; $E$ – electrical intensity, [V/m]; $q$ – charge, [C]; $t$ - time, [seconds]; $\rho$ – specific resistance of insulator, [Ohm-meter, $\Omega \cdot$m]; $a$ – internal radius of the ball, [m]; $C$ – capacity of the ball, [C]; $k = 9\times10^9$ [N$\cdot$m$^2$/C$^2$, m/F]. Last equation is result.

*Example:* Let us take typical data: $\rho = 10^{19}$ $\Omega$-m, $k = 9\times10^9$, $\delta/a = 0.2$, then $t_h = 1.24\times10^7$ seconds = 144 days.

(b) *Half-life of cylindrical tube.* The computation is same as for tubes (1 m charged cylindrical condenser):

$$q = q_0 \exp\left(-\frac{1}{RC}t\right), \quad C = \frac{\varepsilon}{k\ln(1+\delta/a)}, \quad R = \frac{\rho\delta}{2\pi a}, \quad -0.693 = -\frac{1}{RC}t_h,$$

$$t_h = \frac{0.693\rho\delta\varepsilon}{2\pi ka\ln(1+\delta/a)}, \quad for \quad \delta \to 0, \quad final \quad t_h \approx 0.7\frac{\rho\varepsilon}{2\pi k}.$$

(14)

*Example:* Let us take typical data (polystyrene) : $\rho = 10^{18}$ $\Omega\cdot$m, $k = 9\times10^9$, $\varepsilon = 2$, then $t_h = 2.5\times10^7$ seconds = 290 days.

**10. Condenser as accumulator of launch energy.**

Space launcher needs in much energy in sort time. Most researchers of the electromagnetic launcher offer condensers for storage of energy. Let us estimate the maximum energy which can be accumulated by a 1 kg of a plate electric condenser.

$$W_M = \frac{1}{2}Q_M U = \frac{\varepsilon E_s^2}{8\pi k\gamma}, \quad where \quad Q_M = \frac{Q}{M} = \frac{\varepsilon E_s}{4\pi k\gamma d}, \quad C_M = \frac{Q_M}{U} = \frac{\varepsilon}{4\pi k\gamma d^2},$$

(15)

where $W_M$ is energy [J/kg], $Q_M$ is electric charge [C/kg], $U$ is voltage [V], $C_M$ is value of capacitor [C/kg], $\gamma$ is specific density of dielectric [kg/m$^3$], $d$ is distance between plate (layers) in plate condenser [m].

For $\varepsilon = 3$, $E_s = 3\times10^8$ V/m, $k = 9\times10^9$ we have $W_M = 660$ J/kg. The industry capacities have energy density $0.02 \div 0.08$ Wh/kg, the ultracapacity has $3 \div 5$ Wh/kg (1Wh = 3600 J). That is very small value. The energy of a battery is $30 \div 40$ Wh/kg, a gunpowder is about 3 MJ/kg, the energy of a rocket fuel is 9 MJ/kg (C + O$_2$ = CO$_2$). In previous works (see, for example, [25],[40], [60]) the author offered to use as energy accumulator the fly-wheel. The fly-wheel energy storages is

$$W_M = \frac{1}{2}\frac{\sigma}{\gamma},$$

(16)

where $\sigma$ is safety tensile stress [N/m$^2$] of fly-wheel material. For $\sigma = 300$ kg/mm$^2$, $\gamma = 1800$ kg/m$^3$ (it is current composite matter from artificial fibers) we have $W_M = 0.83\times10^5$ J/kg. When we have available composite matter composed of whiskers and nanotubes that critical value will increase many times.

The other method is getting a high electric energy from impulse magneto-dynamic electric generator.

# Projects

Below the reader find the estimation of two projects: Climber for space elevator and Earth space AB launcher. The taken parameters are not optimal. Our aim to illustrate possibilities of offered systems and method computation.

### 1. Space elevator climber.

Let us take the following data: $a = 0.05$ m, $E = 10^8$ V/m, $l = 0.3$ m, $h = d = 3a$, $\varepsilon = 3$. $k = 9 \times 10^9$, $b = 0.109$.

Then the trust of two cylinder electrostatic engine (fig.2a,b,c) is [eq. (4)]

$$F = 2b\frac{\varepsilon E^2 al}{k} = 10.8 \times 10^3 \ N \ .$$

The charge of one cylinder is [eq. (6)]

$$q = \frac{\varepsilon a^2 El}{k} = 2.5 \times 10^{-5} \ C \ .$$

Requested voltage of electric line is $U = Ed = 10^8 \times 0.15 = 15 \times 10^6$ V, [eq.(7)].
Requested maximum power for $V = 0.5$ km/s is $P = FV = 5.4 \times 10^7$ W, [eq.(8)].
Requested currency in electric line is $I = P/U = 3.6$ A, [eq.(9)].
Maximum loss of voltage in electric line for double aluminum wire having length $L = 36,000$ km (up Geosynchronous Earth Orbit), cross-section areas $s = 1$ mm$^2$ ($\rho = 2.8 \times 10^{-6}$ $\Omega$·cm) and coefficient efficiency of Earth electric line [eq.(10)].

$$R = \rho\frac{L}{s} = 2.02 \times 10^6 \ \Omega, \quad \Delta U = IR = 7.27 \times 10^6 \ V, \quad \eta = \frac{U - \Delta U}{U} = \frac{15 \times 10^6 - 7.27 \times 10^6}{15 \times 10^6} = 0.52 \ .$$

As you see the cross-section of wire may be small and coefficient efficiency is good for this super long electric line. We take a maximum length to geosynchronous orbit. That is efficiency delivery energy to GEO station, where a weight of climber is zero. In reality, if we take an average distance and the average climber weight, the loss of energy decreases in 5 times and an efficiency reaches $0.8 \div 0.9$.

The maximum number of main switching is $\upsilon \approx 1.06 \times 10^5$ 1/s [eq. (11)].

### 2. Earth electrostatic AB space launcher

Let us take the following data: $a = 0.1$ m, $E = 10^8$ V/m, $l = 1$ m, $h = d = 3a$, $\varepsilon = 3$, $N = 4$ (N is number of pair driver cylinders), $b = 0.109$.

Then the trust of the eight cylinder electrostatic engine (fig.2d,e) is [eq. (4)]

$$F = 2 \times 4 \times b\frac{\varepsilon E^2 al}{k} = 29 \times 10^4 \ N \ .$$

The charge of one cylinder is [eq. (6)]

$$q = \frac{\varepsilon a^2 El}{k} = 3.3 \times 10^{-2} \ C \ .$$

Requested voltage of the electric line is $U = Ed = 10^8 \times 0.3 = 30 \times 10^6$ V, [eq.(7)].
Requested maximum power for $V = 8$ km/s is $P = FV = 2.32 \times 10^9$ W, [eq.(8)].
Requested currency in the electric line is $I = P/U = 77.3$ A, [eq.(9)].
Look your attention the thrust $F = 29$ tons is enough for launch 10 tons space ship (acceleration is $a = 3g$) with conventional people ($100 \div 150$ tourists) if the monorail has length 1100 km. The trained people (cosmonauts) can keep overload 6g and needed length of track is 530 km. The payload which can keep overload 300g needs only the track 11 km [60].

It is using the suggested electrostatic linear engine, we can build a cheap high productive manned (or unmanned) space catapults at present time. This catapult decreases a launch cost up 2 - 4 $/kg and allows to launch thousands tons in year.

That will be simpler then author's cable catapult offered in [23]-[25],[40],[60]. Nanotubes mobile cable and the 109 drive stations are not needed. There is only electrostatic motionless monorail-stator (which produces a running electrostatic wave of charges) and linear permanent charged rotor (driver) connected by cable to a spaceship. The monorail-stator (cable) is suspended on columns (or in air as in [41] - [42]).

Installation may also be used as high speed conventional tramway.

### 3. Other applications and estimations

1. *Electrostatic Interplanetary Space Launcher or Ship Propulsion.* Assume we want to launch mass $M = 2$ kg interplanetary probe by $L = 100$ m electrostatic accelerator (launcher). We use a thrust $F = 120$ tons. Then the acceleration will be $a = F/M = 1.2 \times 10^6/2 = 0.6 \times 10^6$ m/s$^2$, final speed $V = (2 \times 10^2 \times 0.6 \times 10^6)^{0.5} = 11$ km/s. If we launch from space ship, the space ship receives the momentum $MV$ in opposed direction.

3. *Transport cable systems (space climber) for Earth to Moon, Earth to Mars.* In [25], [29], [30], [36], [39], [60] the author offered and researched the mechanical cable transport systems for Space Elevator and for Earth-Moon, Earth-Mars trips. All these systems need in high speed engine for moving of space vehicle. One (cable) version of the suggested transport system is noted in above cited works. However, the system offered in given article may be used in many structures. The cable is stator, the vehicle has linear rotor. The cable delivers the energy in form of running charge wave, the vehicle (climber) follows this running wave. The speed of running wave (vehicle) can be very high. The voltage is extremely big and the weight of the electric wires is small.

Let us to make the simplest estimation. Assume, the climber weights $W = 1$ ton $= 10,000$ N and has speed $V = 1$ km/s $= 1000$ m/s. The power is $P = WV = 10^4 \times 10^3 = 10^7$ W. For voltage $U = 10^8$ V, the electric currency is $i = 10^7/10^8 = 0.1$ A. For safety currency 20 A/mm$^2$, the need wire diameter is about 0.1 mm$^2$.

4. *Suspended satellite system.* In [25], [36] the author suggested a cable ring rotating around the Earth with motionless satellites suspended within the ring (fig.1e). The offered linear engine can be used as engine for compensation the air friction of cable and as non-contact bearing for suspend system.

5. *Electrostatic levitation train and linear engine.* In [59] the author suggested the electrostatic levitation train (fig.1h). The offered linear engine can be used as propulsion engine for this train. In braking the energy of acceleration will be returned in electric line.

6. *Electrostatic rotary engine.* At present time industry uses conventional low voltage electric engine. When we have a high voltage electric line it may be easier to use high voltage electrostatic rotary engine.

7. *Electrostatic levitation bearing.* Some technical installations need low friction bearings. The mono-electrets can be used as non-contact bearing having zero mechanical friction.

8. *Electrostatic Gun System.* Cannonry needs high speed shells. However, the shell speed is limited by gas speed into cannon. The suggested linear electrostatic engine can be used as the high efficiency shells in armor-piercing cannonry having very high initial shell speed because the initial shell speed of linear engine does not have the speed limit (see application 1 above). That means the electrostatic gun can shoot thousands of kilometers.

### Conclusion

The suggested space climber is single reality high efficiency transport system for space elevator at present time. The electromagnetic beam transfer energy is very complex, expensive and has very low efficiency especially for long distance (from divergence of electromagnetic beam). The laser has same disadvantages. The conventional electric line with conventional electric motor is very heavy and not acceptable for space.

The offered electrostatic engine could find wide application in many fields of technology. That can decrease the monetary costs of launch cost by hundreds to thousands of times. The electrostatic engine needs a very high-voltage but this voltage is located in small area inside of installations and is not dangerous to people. Currently used technology does not have another way for reaching a high speed except by the use of rockets. But rockets and rocket launches are very expensive and we do not know ways to decrease the cost of rocket launch by hundreds to thousands of times.

### *Acknowledgement*

The author wishes to acknowledge Richard Cathcart for correcting the English and offer useful advice.

### References

(Part of these articles the reader can find in author WEB page: http://Bolonkin.narod.ru/p65.htm ; http://arxiv.org , search "Bolonkin"; and in the books: "*Non-Rocket Space Launch and Flight*", Elsevier, London, 2006, 488 pgs., "*New Concepts, Ideas and Innovations in Aerospace and Technology*", Nova, 2007.)

1  Bolonkin, A.A., (1965a), "Theory of Flight Vehicles with Control Radial Force". Collection *Researches of Flight Dynamics*, Mashinostroenie Publisher, Moscow,  pp. 79 -118, 1965, (in Russian). Intern.Aerospace Abstract A66-23338# (English).
2  Bolonkin A.A., (1965c), Optimization of Trajectories of Multistage Rockets. Collection *Researches of Flight Dynamics*. Mashinostroenie Publisher, Moscow, 1965, p. 20 -78 (in Russian). International Aerospace Abstract A66-23337#  (English).
3  Bolonkin, A.A., (1982a), Installation for Open Electrostatic Field, Russian patent application #3467270/21 116676, 9 July, 1982 (in Russian), Russian PTO.
4  Bolonkin, A.A., (1982b), Radioisotope Propulsion. Russian patent application #3467762/25  116952, 9 July 1982 (in Russian), Russian PTO.
5  Bolonkin, A.A., (1982c), Radioisotope Electric Generator. Russian patent application #3469511/25 116927. 9 July 1982 (in Russian), Russian PTO.
6  Bolonkin, A.A., (1983a), Space Propulsion Using Solar Wing and Installation for It, Russian patent application #3635955/23 126453, 19 August, 1983 (in Russian), Russian PTO.
7  Bolonkin, A.A., (1983b), Getting of Electric Energy from Space and Installation for It, Russian patent application #3638699/25 126303, 19 August, 1983 (in Russian), Russian PTO.
8  Bolonkin, A.A., (1983c), Protection from Charged Particles in Space and Installation for It, Russian patent application #3644168  136270, 23 September 1983, (in Russian), Russian PTO.
9  Bolonkin, A. A., (1983d), Method of Transformation of Plasma Energy in Electric Current and Installation for It. Russian patent application #3647344  136681 of 27 July 1983 (in Russian), Russian PTO.
10 Bolonkin, A. A., (1983e), Method of Propulsion using Radioisotope Energy and Installation for It.  of Plasma Energy in Electric Current and Installation for it. Russian patent application #3601164/25 086973  of 6 June, 1983 (in Russian), Russian PTO.
11 Bolonkin, A. A.,(1983f),  Transformation of Energy of Rarefaction Plasma in Electric Current and Installation for it. Russian patent application #3663911/25  159775, 23 November 1983 (in Russian), Russian PTO.
12 Bolonkin, A. A., (1983g), Method of a Keeping of a Neutral Plasma and Installation for it. Russian patent application #3600272/25  086993, 6 June 1983 (in Russian), Russian PTO.
13 Bolonkin, A.A., (1983h),  Radioisotope Electric Generator. Russian patent application #3620051/25 108943, 13 July 1983 (in Russian), Russian PTO.

14 Bolonkin, A.A., (1983i), Method of Energy Transformation of Radioisotope Matter in Electricity and Installation for it. Russian patent application #3647343/25 136692, 27 July 1983 (in Russian), Russian PTO.

15 Bolonkin, A.A., (1983j), Method of stretching of thin film. Russian patent application #3646689/10 138085, 28 September 1983 (in Russian), Russian PTO.

16 Bolonkin, A.A., (1987), "New Way of Thrust and Generation of Electrical Energy in Space". Report ESTI, 1987, (Soviet Classified Projects).

17 Bolonkin, A.A., (1990), "Aviation, Motor and Space Designs", Collection *Emerging Technology in the Soviet Union*, 1990, Delphic Ass., Inc., pp.32-80 (in English).

18 Bolonkin, A.A., (1991), *The Development of Soviet Rocket Engines*, 1991, Delphic Ass.Inc.,122 ps. Washington, (in English).

19 Bolonkin, A.A., (1992a), "A Space Motor Using Solar Wind Energy (Magnetic Particle Sail)". The World Space Congress, Washington, DC, USA, 28 Aug. - 5 Sept., 1992, IAF-0615.

20 Bolonkin, A.A., (1992b), "Space Electric Generator, run by Solar Wing". The World Space Congress, Washington, DC, USA, 28 Aug. -5 Sept. 1992, IAF-92-0604.

21 Bolonkin, A.A., (1992c), "Simple Space Nuclear Reactor Motors and Electric Generators Running on Radioactive Substances", The World Space Congress, Washington, DC, USA, 28 Aug. - 5 Sept., 1992, IAF-92-0573.

22 Bolonkin, A.A. (1994), "The Simplest Space Electric Generator and Motor with Control Energy and Thrust", 45th International Astronautical Congress, Jerusalem, Israel, 9-14 Oct., 1994, IAF-94-R.1.368 .

23 Bolonkin, A.A., (2002a), "Non-Rocket Space Rope Launcher for People", IAC-02-V.P.06, 53rd International Astronautical Congress, The World Space Congress - 2002, 10-19 Oct 2002, Houston, Texas, USA.

24 Bolonkin, A.A,(2002b), "Non-Rocket Missile Rope Launcher", IAC-02-IAA.S.P.14, 53rd International Astronautical Congress, The World Space Congress - 2002, 10-19 Oct 2002, Houston, Texas, USA.

25 Bolonkin, A.A.,(2002c), "Inexpensive Cable Space Launcher of High Capability", IAC-02-V.P.07, 53rd International Astronautical Congress, The World Space Congress - 2002, 10-19 Oct 2002, Houston, Texas, USA.

26 Bolonkin, A.A.,(2002d), "Hypersonic Launch System of Capability up 500 tons per day and Delivery Cost $1 per Lb". IAC-02-S.P.15, 53rd International Astronautical Congress, The World Space Congress - 2002, 10-19 Oct 2002, Houston, Texas, USA.

27 Bolonkin, A.A.,(2002e), "Employment Asteroids for Movement of Space Ship and Probes". IAC-02-S.6.04, 53rd International Astronautical Congress, The World Space Congress - 2002, 10-19 Oct 2002, Houston, Texas, USA.

28 Bolonkin, A.A., (2002f), "Optimal Inflatable Space Towers of High Height". COSPAR-02 C1.1-0035-02, 34th Scientific Assembly of the Committee on Space Research (COSPAR), The World Space Congress - 2002, 10-19 Oct 2002, Houston, Texas, USA.

29 Bolonkin, A.A., (2002g), "Non-Rocket Earth-Moon Transport System", COSPAR-02 B0.3-F3.3-0032-02, 02-A-02226, 34th Scientific Assembly of the Committee on Space Research (COSPAR), The World Space Congress - 2002, 10-19 Oct 2002, Houston, Texas, USA.

30 Bolonkin, A. A.,(2002h) "Non-Rocket Earth-Mars Transport System", COSPAR-02 B0.4-C3.4-0036-02, 34th Scientific Assembly of the Committee on Space Research (COSPAR), The World Space Congress - 2002, 10-19 Oct 2002, Houston, Texas, USA.

31 Bolonkin, A.A.,(2002i). "Transport System for Delivery Tourists at Altitude 140 km". IAC-02-IAA.1.3.03, 53rd International Astronautical Congress, The World Space Congress - 2002, 10-19 Oct. 2002, Houston, Texas, USA.

32 Bolonkin, A.A., (2002j), "Hypersonic Gas-Rocket Launch System." AIAA-2002-3927, 38th AIAA/ASME/SAE/ASEE Joint Propulsion Conference and Exhibit, 7-10 July 2002. Indianapolis, IN, USA.

33 Bolonkin, A.A., (2003a), "Air Cable Transport", *Journal of Aircraft*, Vol. 40, No. 2, March-April 2003.

34 Bolonkin, A.A., (2003b), "Optimal Inflatable Space Towers with 3-100 km Height", *JBIS*, Vol. 56, No 3/4, pp. 87-97, 2003.

35 Bolonkin, A.A.,(2003c), "Asteroids as Propulsion Systems of Space Ships", *JBIS*, Vol. 56, No 3/4, pp.

97-107, 2003.

36  Bolonkin A.A., (2003d), "Non-Rocket Transportation System for Space Travel", *JBIS*, Vol. 56, No 7/8, pp. 231-249, 2003.

37  Bolonkin A.A., (2003e), "Hypersonic Space Launcher of High Capability", *Actual problems of aviation and aerospace systems*, Kazan, No. 1(15), Vol. 8, 2003, pp. 45-58.

38  Bolonkin A.A., (2003f), "Centrifugal Keeper for Space Stations and Satellites", *JBIS*, Vol. 56, No 9/10, pp. 314-327, 2003.

39  Bolonkin A.A., (2003g), "Non-Rocket Earth-Moon Transport System", *Advances in Space Research*, Vol. 31/11, pp. 2485-2490, 2003, Elsevier.

40  Bolonkin A.A., (2003h), "Earth Accelerator for Space Ships and Missiles". *JBIS*, Vol. 56, No. 11/12, 2003, pp. 394-404.

41  Bolonkin A.A., (2003i), "Air Cable Transport and Bridges", TN 7567, International Air & Space Symposium - The Next 100 Years, 14-17 July 2003, Dayton, Ohio, USA.

42  Bolonkin, A.A., (2003j), "Air Cable Transport System", *Journal of Aircraft*, Vol. 40, No. 2, March-April 2003, pp. 265-269.

43  Bolonkin A.A.,(2004a), "Kinetic Space Towers and Launchers ', *JBIS*, Vol. 57, No 1/2, pp. 33-39, 2004.

44  Bolonkin A.A.,(2004b), "Optimal trajectory of air vehicles", *Aircraft Engineering and Space Technology*, Vol. 76, No. 2, 2004, pp. 193-214.

45  Bolonkin A.A., (2004c), "Long Distance Transfer of Mechanical Energy", International Energy Conversion Engineering Conference at Providence RI, Aug. 16-19, 2004, AIAA-2004-5660.

46  Bolonkin A.A., (2004d), "Light Multi-Reflex Engine", Journal *JBIS*, Vol. 57, No 9/10, pp. 353-359, 2004.

47  Bolonkin, A.A., (2004e), "Kinetic Space Towers and Launchers", Journal *JBIS*, Vol. 57, No 1/2, pp. 33-39, 2004.

48  Bolonkin, A.A., (2004f), "Optimal trajectory of air and space vehicles", *AEAT*, No 2, pp. 193-214, 2004.

49  Bolonkin, A.A.,(2004g), "Hypersonic Gas-Rocket Launcher of High Capacity", Journal *JBIS*, Vol. 57, No 5/6, pp. 167-172, 2004.

50  Bolonkin, A.A., (2004h), "High Efficiency Transfer of Mechanical Energy". International Energy Conversion Engineering Conference at Providence RI, USA. 16-19 August, 2004, AIAA-2004-5660.

51  Bolonkin, A.A., (2004i), "Multi-Reflex Propulsion System for Space and Air Vehicles", *JBIS*, Vol. 57, No 11/12, 2004, pp. 379-390.

52  Bolonkin A.A.,(2005a) "High Speed Catapult Aviation", AIAA-2005-6221, Atmospheric Flight Mechanic Conference - 2005, 15-18 August, 2005, USA.

53  Bolonkin A.A., (2005a), Electrostatic Solar Wind Propulsion System, AIAA-2005-3653. 41-st Propulsion Conference, 10-12 July, 2005, Tucson, Arizona, USA.

54  Bolonkin A.A., (2005b), Electrostatic Utilization of Asteroids for Space Flight, AIAA-2005-4032. 41 Propulsion Conference, 10-12 July, 2005, Tucson, Arizona, USA.

55  Bolonkin A.A., (2005c), Kinetic Anti-Gravitator, AIAA-2005-4504. 41-st Propulsion Conference, 10-12 July, 2005, Tucson, Arizona, USA.

56  Bolonkin A.A., (2005d), Sling Rotary Space Launcher, AIAA-2005-4035. 41-st Propulsion Conference, 10-12 July, 2005, Tucson, Arizona, USA.

57  Bolonkin A.A., (2005e), Radioisotope Space Sail and Electric Generator, AIAA-2005-4225. 41-st Propulsion Conference, 10-12 July, 2005, Tucson, Arizona, USA.

58  Bolonkin A.A., (2005f), Guided Solar Sail and Electric Generator, AIAA-2005-3857. 41-st Propulsion Conference, 10-12 July, 2005, Tucson, Arizona, USA.

59  Bolonkin A.A., (2005g), Problems of Electrostatic Levitation and Artificial Gravity, AIAA-2005-4465. 41 Propulsion Conference, 10-12 July, 2005, Tucson, Arizona, USA.

60  Bolonkin A.A., (2006a), "*Non-Rocket Space Launch and Flight*", Elsevier, London, 2006, 488 pgs.

61  Bolonkin A.A., (2006b), Electrostatic AB-Ramjet Space Propulsion, AIAA-2006-6173, AEAT, Vol.79, No. 1, 2007, pp. 3 - 16.

62  Bolonkin A.A., (2006c), Beam Space Propulsion, AIAA-2006-7492, (published in http://arxiv.org search "Bolonkin").

63  Bolonkin A.A., (2006d), High Speed Solar Sail, AIAA-2006-4806, (published in http://arxiv.org search

"Bolonkin").

64  Bolonkin A.A., (2006e), Suspended Air Surveillance System, AIAA-2006-6511, (published in http://arxiv.org search "Bolonkin").

65  Bolonkin A.A., (2006f), Optimal Solid Space Tower (Mast), (published in http://arxiv.org search "Bolonkin").

66  Bolonkin A.A., (2006g), Electrostatic Linear Engine, presented as paper AIAA-2006-5229 to 42nd Propulsion Conference, USA and published in AEAT, Vol.78, No.6, 2006, pp. 502-508.

67  Bolonkin A.A., (2006h). AB Levitator and Electricity Storage. http://arxiv.org , search "Bolonkin".

68  Bolonkin A.A., (2006i), Theory of Space Magnetic Sail Some Common Mistakes and Electrostatic MagSail. Presented as paper AIAA-2006-8148 to 14-th Space Planes and Hypersonic System Conference, 6-9 November 2006, Australia. http://arxiv.org , search "Bolonkin".

69  Bolonkin A.A., (2006j) Micro - Thermonuclear AB-Reactors for Aerospace. Presented as paper AIAA-2006-8104 in 14th Space Plane and Hypersonic Systems Conference, 6-8 November, 2006, USA. http://arxiv.org , search "Bolonkin".

70  Bolonkin A.A. (2006k). Simplest AB-Thermonuclear Space Propulsion and Electric Generator. http://arxiv.org , search "Bolonkin".

71   Bolonkin A.A., (2006 *l*). Wireless Transfer of Electricity in Outer Space. AIAA-2007-0590, http://arxiv.org , search "Bolonkin".

72  Bolonkin A.A., New Concepts, Ideas and Innovations in Aerospace and Technology, Nova, 2007.

73  Book (2006): *Macro-Engineering - A challenge for the future*. Collection of articles. Eds. V. Badescu, R. Cathcart and R. Schuiling, Springer, 2006. (Collection contains two Bolonkin's articles: Space Towers; Cable Anti-Gravitator, Electrostatic Levitation and Artificial Gravity).

74  Kikoin, I.K., (ed.), Tables of Physical Values. Atomuzdat, Moscow, 1976 (in Russian).